\documentclass[a4paper,11pt]{article}

\usepackage{pos}
\usepackage{lipsum} %package to generate placeholder text in the following
\usepackage{color}
\usepackage{setspace}
\usepackage{titlesec}
\titlespacing{\section}{0pt}{*1.0}{*0.67}
\titlespacing{\subsection}{0pt}{*1.0}{*0.66}

\usepackage{lineno}
\title{First Array-Wide Search for Diffuse UHE Neutrinos with the Askaryan Radio Array}

\ShortTitle{First Array-Wide Search for Diffuse UHE Neutrinos with ARA}

\author*[a,b,c,d]{Marco Stein Muzio} 
\onbehalf{for the ARA Collaboration \\{\normalsize \normalfont(a complete list of authors can be found at the end of the proceedings)}\\}

\affiliation[a]{Department of Physics, Wisconsin IceCube Particle Astrophysics Center, University of Wisconsin,\\
Madison, WI, USA}
\affiliation[b]{Department of Physics, Pennsylvania State University,\\ University Park, PA 16802, USA}
\affiliation[c]{Department of Astronomy and Astrophysics, Pennsylvania State University,\\ University Park, PA 16802, USA}
\affiliation[d]{Institute of Gravitation and the Cosmos, Center for Multi-Messenger Astrophysics, Pennsylvania State University,\\ University Park,
PA 16802, USA}

\emailAdd{muzio@wisc.edu}

\abstract{

The Askaryan Radio Array (ARA) is an ultrahigh energy (UHE) neutrino detector at the South Pole, designed to search for radio pulses emitted by neutrino-initiated particle showers in ice. ARA consists of an array of five autonomous stations with 2 km spacing. Each station consists of 16 radio antennas embedded ${\sim}200$~m deep in the ice that are sensitive to either vertically- or horizontally-polarized signals. Radio arrays like ARA represent a cost-efficient means of achieving the enormous detection $\mathcal{O}(10~\text{km}^3)$ volumes necessary for UHE neutrino detection. This contribution presents the current status of the first-ever array-wide search for UHE neutrinos, leveraging ARA’s unprecedented ${\sim}28$~station-years of livetime. This search will have the best sensitivity of any neutrino detector above $3$~EeV, sufficient to probe the $220$~PeV flux inferred from KM3NeT's observation of KM3-230213A. Importantly, this study demonstrates the feasibility of array-wide neutrino searches, which are necessary for next-generation detectors, like RNO-G (35~stations planned) and IceCube-Gen2 Radio (361~stations proposed), to achieve their design sensitivity. We discuss the progress towards a fully analyzed sample and improvements to ARA’s detector characterization and analysis sensitivity.
}

\ConferenceLogo{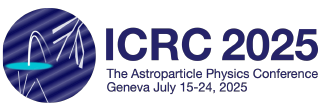}

\FullConference{39th International Cosmic Ray Conference (ICRC2025)\\
 15–24 July 2025\\
Geneva, Switzerland\\}

\begin{document}

\maketitle

\section{Introduction}\label{sec:intro}
% UHE neutrino expectation
% UHE neutrino production channels 
% UHE neutrino science potential
% observational needs

\par
The diffuse flux of ultrahigh energy (UHE) neutrinos ($E_\nu \gtrsim 10^{16.5}$~eV) is beginning to emerge~\cite{KM3NeT:2025npi}, having long been predicted as a consequence of UHE cosmic ray (CR, $E_\mathrm{CR} \gtrsim 10^{18}$~eV) interactions~\cite{Greisen:1966jv,Zatsepin:1966jv}. UHECRs interact photohadronically with the cosmic microwave background (CMB), extragalactic background lights (EBL), and source ambient photon fields. These interactions produce pions via the $\Delta$-resonance, which (for $\pi^+$) subsequently decay to $\mu^+ + \nu_\mu \rightarrow e^+ + \bar{\nu}_\mu + \nu_\mu + \nu_e$. These neutrinos carry roughly $1/20$ of the CR energy-per-nucleon. CRs also interact hadronically in their source environment, but the subsequent neutrinos have a much smaller energy. These are the only Standard Model production channels of UHE neutrinos, making their observation a smoking gun signature of UHECR interactions. Moreover, UHE neutrinos can probe UHECR sources and their properties on cosmological scales, whereas UHECRs themselves can only probe sources within ${\sim} 100$~Mpc. For these reasons, observation of UHE neutrinos would provide an unprecedented window into the Universe's most extreme astrophysical environments.

\par
Empirically, the UHE neutrino flux is exceedingly small with current constraints limiting the flux $\lesssim 10^{-8}$~GeV/cm$^2$/s/sr at $1$~EeV~\cite{IceCube:2025ezc}. This necessitates detectors with $\mathcal{O}(10~\text{km}^3)$~volumes. Fortuitously, in dense dielectric media UHE neutrino interactions initiate a particle shower which emits a sub-nanosecond pulse of coherent radio emission along the Cherenkov cone, known as Askaryan emission~\cite{Askaryan:1961pfb,ANITA:2006nif}. In glacial ice, the $\mathcal{O}(\text{km})$ attenuation length of radio waves makes observation of these pulses a promising and cost-effective means to detect UHE neutrinos. 

\par
Several experiments have been designed to take advantage of this effect by using sparse arrays of independent radio detector stations, including the Askaryan Radio Array (ARA), ARIANNA~\cite{Barwick:2016mxm}, RNO-G~\cite{RNO-G:2020rmc} (under construction), and IceCube-Gen2 Radio~\cite{IceCubeGen2_TDR} (proposed). Among these, ARA has the largest exposure to date and, in this proceeding, we report on the status of its first array-wide search for diffuse UHE neutrinos.

\section{The Askaryan Radio Array}\label{sec:ara}
% overview of detector
% vanilla station summary
% phased array station summary

\par
\begin{figure}[h!]
  \centering
  \includegraphics[width=0.49\textwidth]{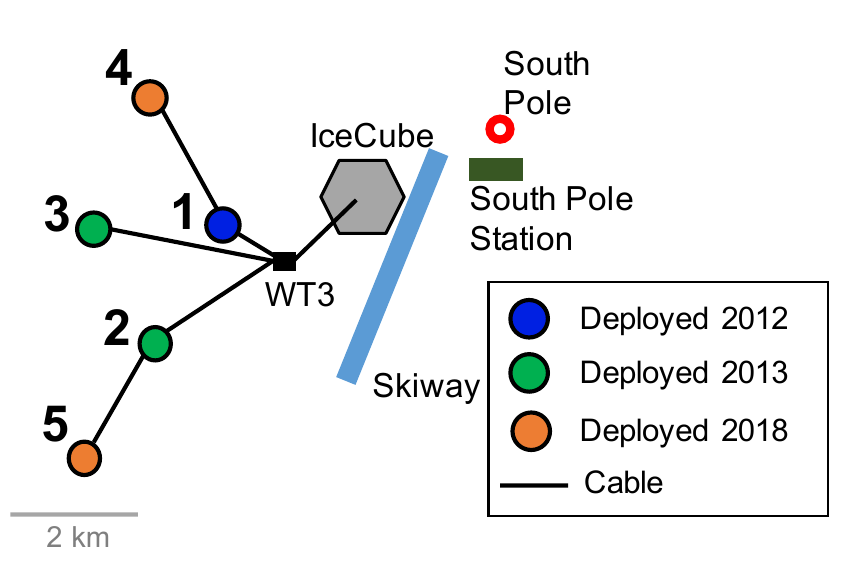}
  \hfill
  \includegraphics[width=0.49\textwidth]{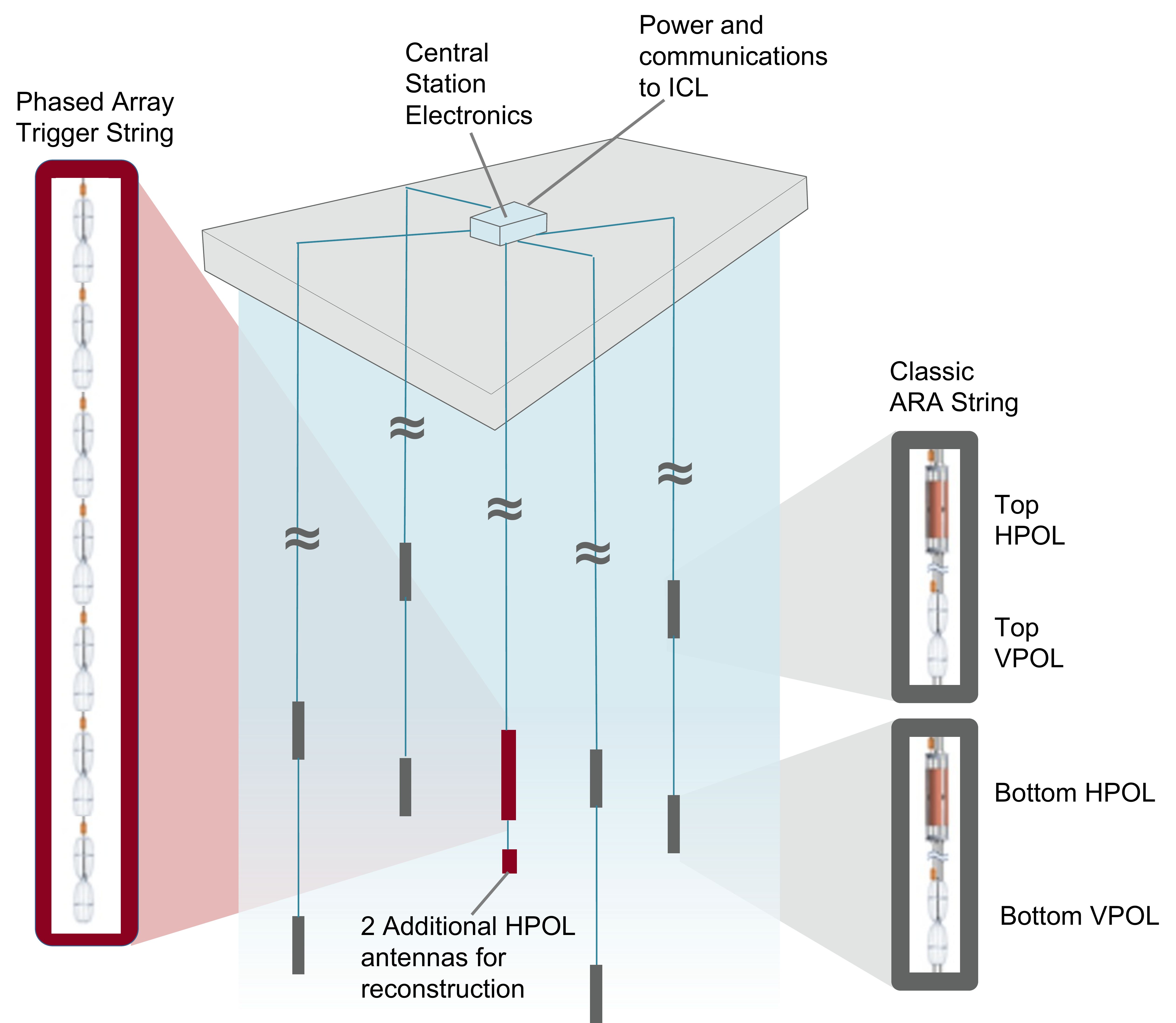}
  \caption{Left: The ARA array, relative to IceCube and the South Pole. Right: Layout of the two subdetectors of A5. Omitting the central PA string (red), other stations of ARA (A1-4) have the same layout.}
  \label{fig:ara_overview}
\end{figure}

The ARA detector comprises $5$~independent stations on a hexagonal grid with ${\sim}2$~km spacing at the South Pole (see left panel of Fig.~\ref{fig:ara_overview}). Each station has $4$~strings deployed up to ${\sim}200$~m deep in the ice. These strings each have $2$~pairs of antennas sensitive to vertically- and horizontally-polarized signals, respectively, in the $150-850$~MHz band. Additional strings are deployed at similar depths with pulsers in each polarization, which are used for calibration and monitoring. This baseline station design was chosen to optimize the effective area to ${\gtrsim}1$~EeV neutrinos. 

\par
ARA stations trigger when an integrated power 5 times larger than the ambient noise level in a $25$~ns window is observed in 3 same-polarization antennas within a $170$~ns coincidence window. This results in a ${\sim}6$~Hz trigger rate, including a $1$~Hz calibration pulser trigger. Additionally, a $1$~Hz forced software trigger is taken to monitor the ambient noise environment of the station.

\par
The fifth station of ARA, A5, has an additional subdetector called the phased array (PA). The PA consists of a central string of $7$~vertically-polarized and $2$~horizontally-polarized antennas. The PA adds signals in each antenna offset by time delays corresponding to pre-defined arrival directions, called beams. Since their close ${\sim}2$~m spacing ensures all antennas see approximately the same signal, a real signal arriving from the direction corresponding to a beam will sum coherently, therefore boosting its signal-to-noise ratio (SNR). Conversely, thermal noise will usually add incoherently. The PA performs this sum before the trigger, making it more efficient to low SNR signals~\cite{Allison:2018ynt} and yielding a dataset with improved signal-background discrimination in analysis~\cite{ARA:2022rwq}. 

\par
The PA triggers when a beam has power in a $10$~ns window above a threshold set to maintain an ${\sim}11$~Hz global trigger rate across all $15$~beams. The traditional ARA and PA systems had separate DAQs previous to $2020$ and triggers on the PA were used to force a trigger on the traditional system. However, at the end of $2019$, the traditional ARA DAQ on A5 was lost due to a USB failure. For this reason, $7$~vertically-polarized antennas from the traditional DAQ were connected to the PA DAQ (limited by the number of open channels). For this reason, from $2020$ onwards, the merged system is often referred to as A5/PA. The station layout for A5 is depicted in the right panel of Fig.~\ref{fig:ara_overview}.

\section{An Array-Wide Search for Diffuse UHE Neutrinos}\label{sec:arrayWideSearch}
% introduce overall analysis and livetime, place in the field
% compare to previous ARA analyses
% mention multi-institution team

\par
ARA has the largest exposure to UHE neutrinos of any in-ice array of radio detectors. In this analysis, we perform the first-ever array-wide search for UHE neutrinos. We consider data taken between $2013$ and $2023$, amounting to approximately $28$~station-years of livetime and ${\sim}379$~TB of data on disk. The analysis is developed and optimized on a ``blinded'' $10\%$ subset of the full data to mitigate biases. After cleaning and optimization is complete, the analysis pipeline and selection cuts are fixed before the full $100\%$ data is processed. In this proceeding, we focus on the development and optimization of the analysis pipeline using the $10\%$ data.

\par
Analysis of this dataset represents a significant challenge, both computationally and organizationally. A highly coordinated $8$-person, multi-institution team has worked to overcome these challenges. This team has made substantial improvements to ARA's existing data access and simulation frameworks, \verb|AraRoot|~\cite{araroot} and \verb|AraSim|~\cite{arasim}, and developed a new analysis framework, \verb|AraProc|~\cite{araproc}. This work has yielded a streamlined and unified analysis of the ARA data across a large, widely dispersed team --- in addition to improving ARA's detector and neutrino simulations. 

\subsection{Unified Analysis Framework}\label{sec:unifiedFramework}
% overview of araproc
% summary of major backgrounds
%% untagged calpulsers
%% cw contamination 
%% pole/station activities
%% cosmic rays
% summary of major cuts
%% CP cut, bad run cut/first minute cut, trigger tag cut, spatiotemporal cut, zenith cut
%% figures: Gumbel fit, CP cut and reconstruction of RF triggers in theta/phi or a skymap

\par
\verb|AraProc| is general purpose tool to provide: 1) a unified data \& simulation processing tool to ensure all waveforms are due to calibrated, interpolated, dedispersed, and filtered in a unified way; 2) a library of functions to perform interferometric vertex reconstruction and calculate summary variables (including SNR, impulsivity, maximum cross-correlation, etc.). A unified library of analysis-specific scripts to perform data processing \& cleaning, investigate events, implement selection cuts, and calculate sensitivities at scale has also been developed by building on this powerful backend. These tools have allowed for the efficient investigation and removal of backgrounds.  

\par
Neutrino signals in ARA's data are expected to be highly impulsive and (most likely) to reconstruct below the station. Given the existing constraints~\cite{ANITA:2019wyx,IceCube:2025ezc}, no more than $\mathcal{O}(10)$ neutrinos are expected to exist in the full dataset (and therefore fewer than $\mathcal{O}(1)$ neutrinos in the $10\%$~sample presented here). Therefore, the ARA data is overwhelmingly dominated by backgrounds. These backgrounds can be categorized into two major categories: impulsive and non-impulsive. 

\par
\begin{figure}[h!]
  \centering
  \includegraphics[width=\textwidth]{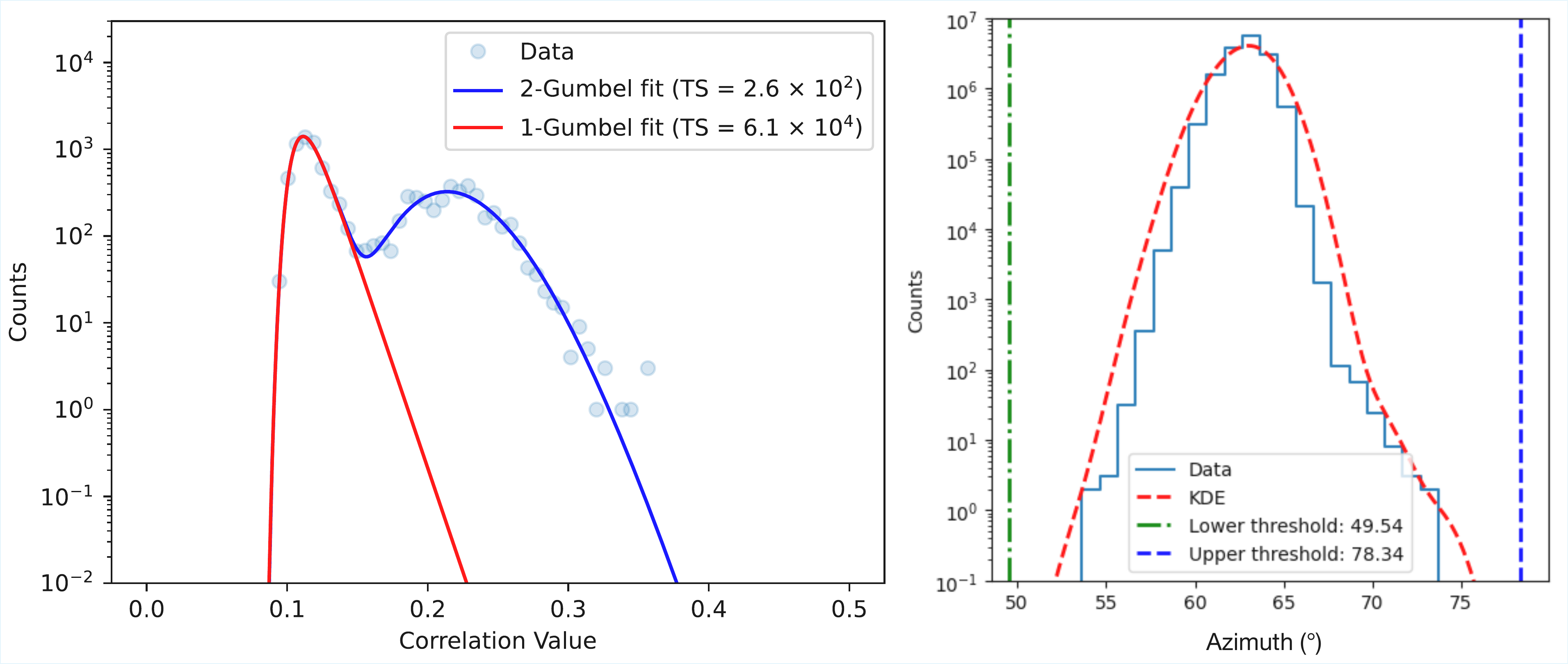}
  \caption{Left: Example of a highly contaminated run which is better described by a two-population model, as evidenced by its lower test statistic (TS). Right: Example of a calibration pulser distribution, along with its KDE fit used to set the geometric cut (vertical lines).}
  \label{fig:background_cuts}
\end{figure}

After basic data quality cuts, the major impulsive backgrounds in the ARA dataset are calibration activity, anthropogenic activity, and cosmic rays. Targeted cuts are used to remove these backgrounds, while still preserving the majority of (simulated) neutrino events. Logged calibration activities are removed by excluding the corresponding data from analysis, while untagged calibration pulses are removed via a geometric cut on events' directional reconstruction (see Fig.~\ref{fig:background_cuts} right panel). Unlogged calibration and other anthropogenic activities are more challenging to remove. Data runs highly contaminated by such activities are systematically identified by analyzing the distribution of correlation values. Uncontaminated runs have a single population of thermal events, whereas contaminated runs have both thermal and non-thermal populations. We remove runs much better described by a two-population hypothesis than a single-population hypothesis (see Fig.~\ref{fig:background_cuts} left). Other unlogged anthropogenic activities are removed via spatiotemporal cluster cuts and additional geometric cuts. 

\par
The most challenging impulsive background to remove is that of cosmic rays. Cosmic rays produce impulsive radio pulses via geomagnetic and Askaryan emission~\cite{Schroder:2016hrv}. This makes their expected signals nearly identical to those expected from neutrinos. While ongoing~\cite{Ali:2025ICRC,Windischhofer:2025ICRC} and future studies may reveal more efficient means of neutrino-CR discrimination, ARA's CR background is removed by cutting all signals that reconstruct above the station, since CRs only come from this region. This results in a roughly $10-30\%$ loss in signal efficiency depending upon the station.

\par 
ARA's data has two primary non-impulsive backgrounds. The first is continuous wave (CW) contamination, primarily from station communications, satellite communications, and weather balloons. To save livetime which is CW contaminated, events are filtered using techniques first developed by the ANITA Collaboration~\cite{ANITA:2018vwl}. The other non-impulsive background, and by far ARA's largest background, is that of thermal events. Such events have properties similar to near-threshold neutrino events. To most efficiently remove these events, all events' summary variables are combined via a linear discriminant to maximize the separation between thermal events and simulated neutrinos. A cut, optimized for the strongest limit on the UHE neutrino flux, is made on the linear discriminant score to remove these thermal backgrounds.

\subsection{Improvements to Detector Simulations}\label{sec:detectorSims}
% data-driven noise models
% antenna gain model improvements
% system gain modeling
%% figure: updated noise models and data-MC agreement

\par
Major improvements have been made to ARA's detector characterization and simulation. First, improved anechoic chamber measurements have been made of ARA's three antenna types: horizontally-polarized antennas, and top \& bottom vertically-polarized antennas, which differ in their number of through-cables. These measurements have provided better characterization of the frequency-dependent gain pattern of the detector's antennas. 

\par
\begin{figure}[h!]
  \centering
  \includegraphics[width=\textwidth]{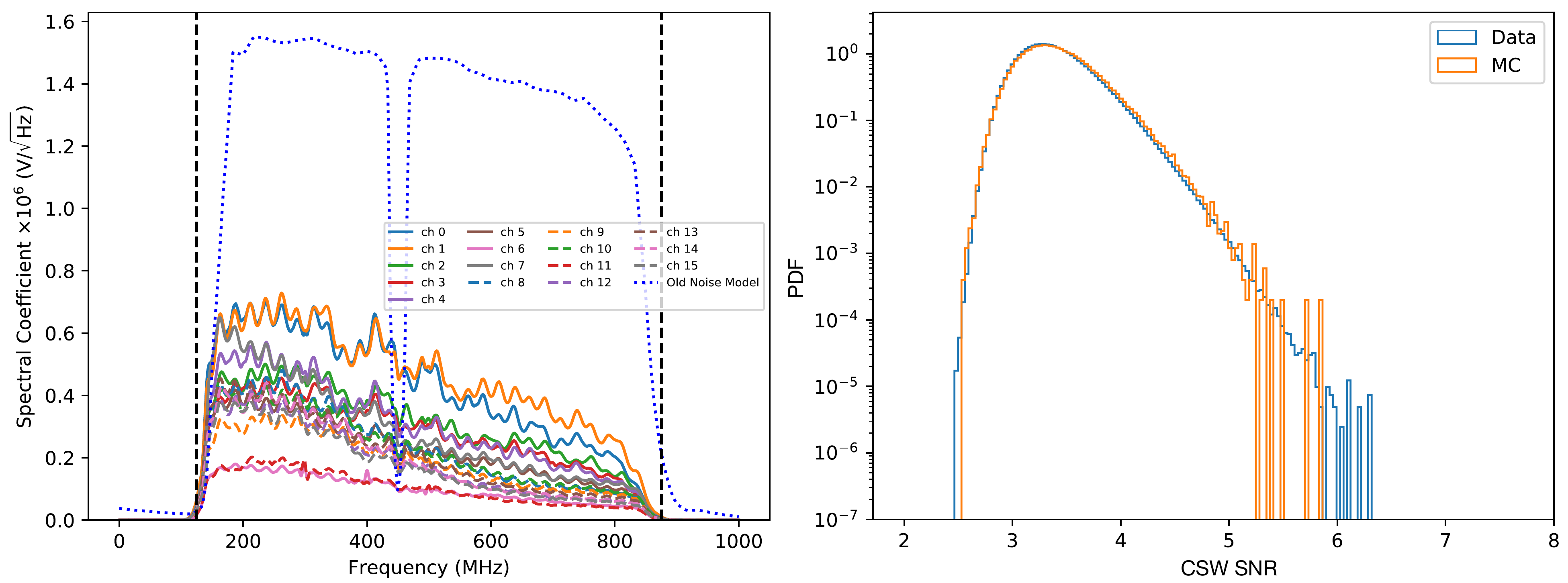}
  \caption{Left: Example per-channel data-driven noise models for A3. The previously used models are shown in blue dotted lines. The ARA frequency band is denoted by dashed vertical lines. Right: Illustration of data-Monte Carlo agreement for the coherently summed waveform (CSW) SNR of about $100$k simulated noise events (orange) and about $10$M real A5/PA data events (blue), which is dominated by thermal noise.}
  \label{fig:data_mc}
\end{figure}

Data-driven noise models have been developed for each station and channel by analyzing the distribution of spectral amplitudes of forced trigger events. These distributions were fit at each frequency to a Rayleigh distribution, and their widths (or spectral coefficients) were recorded. From these models, noise traces are created by generating a random phase and a frequency spectrum with spectral amplitudes drawn from the measured Rayleigh distributions. These data-driven noise models are compared to the previous noise parameterization in Fig.~\ref{fig:data_mc} (left). Further, data-driven signal chain gain models are created for each station and channel by taking the ratio of the observed noise level to the theoretical expectation for thermal noise, given the ice temperature and antenna \& amplifier properties. Noise and gain models are only created for periods called livetime configurations, rather than for each data run individually, since the properties of a station are often stable over long periods. These refinements to ARA's detector simulation have significantly improved our data-Monte Carlo agreement, as illustrated in Fig.~\ref{fig:data_mc} (right).

\subsection{Improvements to Neutrino Simulations}\label{sec:neutrinoSims}
% nulepton sim -> arasim pipeline
% role of secondaries and coincidences
% cite Abby's proceeding
%% figure: array-wide livetime configs 

\par
In addition to general improvements to the ARA simulation framework, \verb|AraSim|, we have completed a significant refactoring to better account for secondary particles and events triggering multiple stations. To accomplish this, an isotropic flux of primary neutrinos uniformly intersecting a $15$~km radius, $3$~km high cylinder centered on A2 is generated by \verb|NuLeptonSim|~\cite{Cummings:2023iuw} and propagated forward to this detection volume. All primary \& secondary interaction vertices within this cylinder are saved and input to single-station \verb|AraSim| simulations, which handle signal generation \& propagation and detector simulation. This results in a set of station-specific simulated traces along with an array-wide list of triggers, and fully accounts both for triggers from secondary particles (which account for up to ${\sim}50\%$~to the effective volume) and for events that trigger multiple stations.

\par
\begin{figure}[h!]
  \centering
  \includegraphics[width=\textwidth]{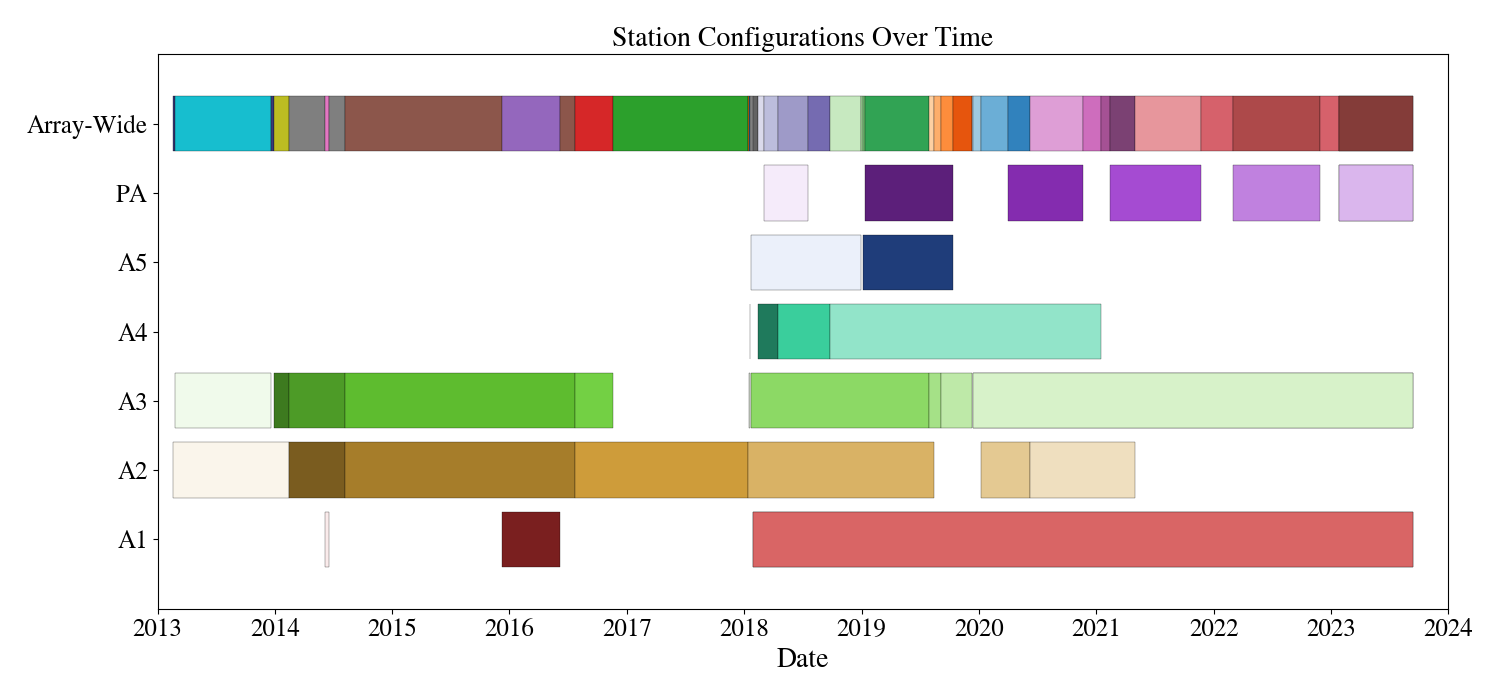}
  \caption{Summary of array-wide and station-level livetime configurations. Each color in a row represents a unique livetime configuration. Colors between rows are unrelated. Station configurations represent differences in detector settings, properties, or noise levels, while array-wide configurations represent changes in station-level configurations.}
  \label{fig:veff_configs}
\end{figure}

This modularized framework allows for the efficient calculation of an array-wide effective volume, which would otherwise be expensive as it requires an array-wide simulation of the detector, since stations cannot be approximated as having independent detection volumes at high energies. In particular, station-level livetime configuration changes can be accounted for without resimulation of the full array. This is advantageous since $2013-2023$ individual stations had as many as $9$~stable livetime configurations. After excluding bad quality runs, this gives rise to $39$~unique array-wide livetime configurations. Figure~\ref{fig:veff_configs} summarizes the changes in livetime configurations of the array over this period and Fig.~\ref{fig:sensitivity} (left) shows the effective volume over one configuration. See~\cite{Bishop:2025ICRC} for a detailed discussion of this effective volume calculation.

\subsection{Array-Wide Optimization \& Sensitivity}\label{sec:arrayWideOptimization}
% summary of optimization plan
%% per-station-config lda
%% array-wide optimization
%% overview of array-wide livetime configurations
% projected limit and event rates
%%% figures: lda preview and projected limit

\par
After all impulsive backgrounds and CW contamination are removed from the $10\%$~data sample, the final step of the analysis is to train a linear discriminant (LD) for each station-livetime configuration and set a cut in the LD score, $t_s$, to remove thermal backgrounds, which represent the overwhelming majority of events. This cut will be set by optimizing array-wide for the strongest limit. The flux limit itself depends upon two quantities: the array-wide signal efficiency $\epsilon(\mathbf{t})$ and the array-wide background leakage rate $\sum_s b_s(t_s)$, where $s$ is the station and $\mathbf{t}=(t_1,t_2,t_3,t_4,t_5)$. Since we want to optimize the limit for the full livetime of the array, we must also perform this optimization over each array-wide livetime configuration of the detector by minimizing

\begin{align}
    \phi_0^\mathrm{UL} = \frac{\mathrm{FC}(\sum_{c,s} b_{c,s}(t_{c,s}))}{4\pi\int dE f(E) \sum_{c,s} T_c A_{\mathrm{eff},c}(E) \epsilon_c(E, \mathbf{t}_c) }~,
\end{align}

\noindent
assuming a neutrino flux of the form $\phi(E) = \phi_0 f(E)$, for a dimensionless function $f(E)$, and where $\mathrm{FC}(x)$ is the Feldman-Cousins $90\%$ confidence level (CL) upper-limit on a background $x$, $c$ is the array-wide livetime configuration, $T_c$ is the livetime, and $A_{\mathrm{eff},c}$ is the array-wide effective area. For optimization, we will assume a neutrino flux given by the global upper-limit, which is a union of the current limits from IceCube~\cite{IceCube:2025ezc} and ANITA~\cite{ANITA:2019wyx}.

\par
Analysis of the full $100\%$ data will yield either candidate UHE neutrino events or the strongest limit on their flux by any in-ice radio experiment to date. Given the large exposure of this analysis, several neutrino candidates may exist in the dataset at trigger level under the most optimistic flux models: $39.3$~events under the global upper-limit model~\cite{IceCube:2025ezc,ANITA:2019wyx}; $27.5$~events under the cosmogenic maximum proton model of~\cite{Muzio:2023skc}; and $5.4$~events under the model of~\cite{Kotera:2010yn}.

\par
\begin{figure}[h!]
  \centering
  \includegraphics[width=0.48\textwidth]{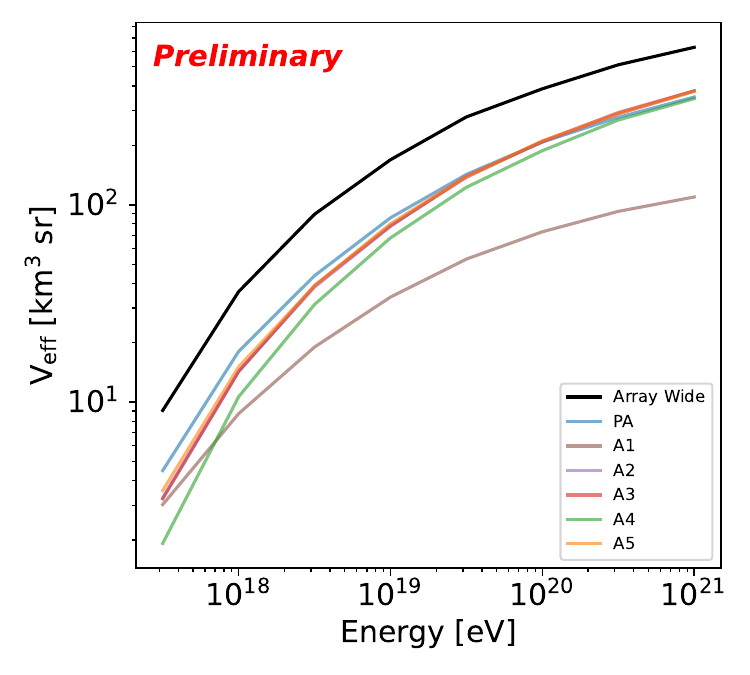}
  \hfill
  \includegraphics[width=0.51\textwidth]{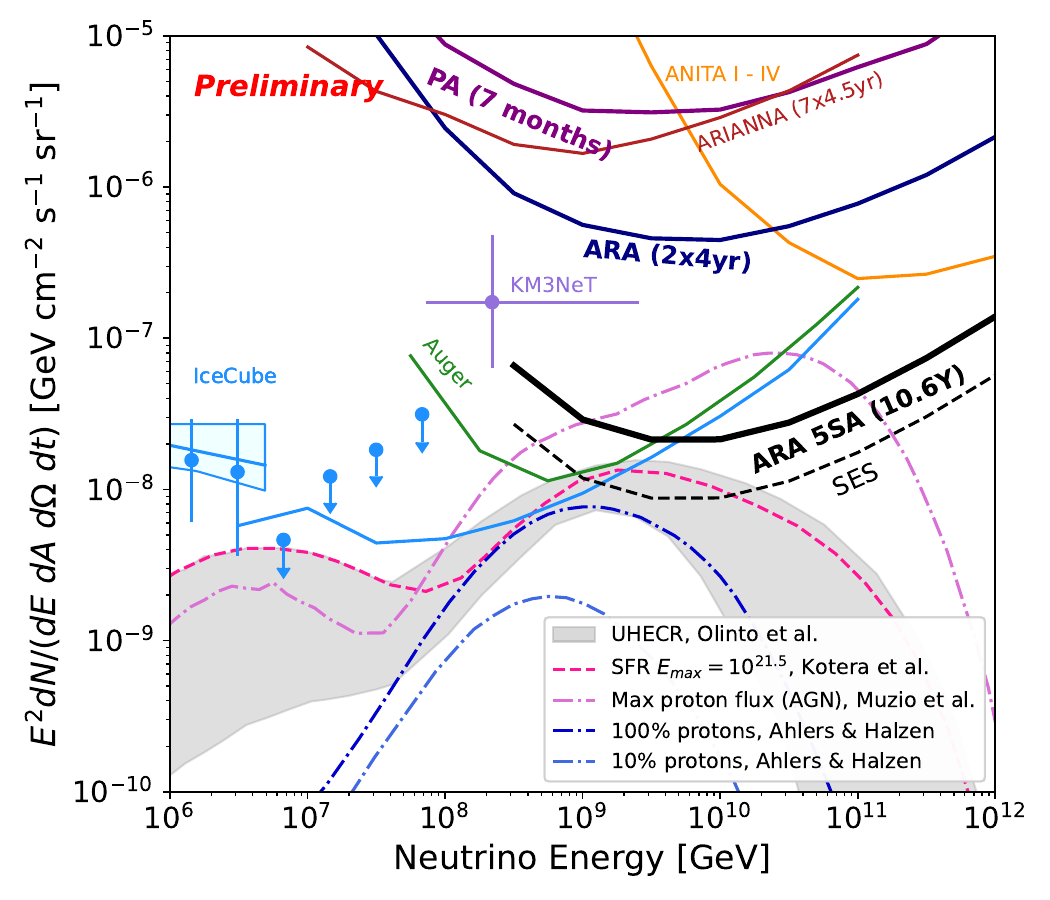}
  \caption{Left: Array-wide effective volume for one array-wide livetime configuration compared with those for single stations. Right: Projected analysis-level sensitivity for this search in terms of the projected $90\%$~confidence level upper limit (solid) and single-event sensitivity (dashed) adopting analysis efficiencies from previous ARA searches~\cite{ARA:2019wcf,ARA:2022rwq}. Data, limits, and model predictions are shown for reference.}
  \label{fig:sensitivity}
\end{figure}

As a conservative sensitivity projection of this analysis, we scale the previous analysis-level limits obtained in~\cite{ARA:2019wcf,ARA:2022rwq} by the approximate exposure of this analysis, accounting for the increased livetime and effective area. The projected $90\%$ CL upper-limit is shown in Fig.~\ref{fig:sensitivity} (right) and would be the strongest limit up to $1000$~EeV from any radio detector experiment.

\section{Conclusion}\label{sec:conclusion}
% summarize analysis
% summarize expected impact

\par
The ARA Collaboration is currently undertaking the first-ever array-wide search for UHE neutrinos using a sparse array of in-ice radio detector stations. A multi-institution team has conducted this search through ${\sim}28$~station-years of livetime, leveraging a new set of unified analysis tools. This pioneering analysis will demonstrate the feasibility of array-wide searches, which are necessary for next-generation experiments, like RNO-G ($35$~stations planned) and IceCube-Gen2 Radio ($361$~stations proposed), to fully realize their planned sensitivity.

\par
Improvements in analysis tools have made more systematic removal of both impulsive and non-impulsive backgrounds possible. At the same time, advances in detector and neutrino simulations have significantly improved data-Monte Carlo agreement and made efficient, accurate calculation of array-wide effective volumes possible. These gains will be fully leveraged by performing the final optimization to remove thermal backgrounds both array-wide and across ARA's entire livetime. 

\par
This analysis has significant potential to observe a UHE neutrino candidate, and is about $3$~times more sensitive than the central flux value inferred from KM3-230213A~\cite{KM3NeT:2025npi}. If not, this analysis will place the strongest UHE neutrino flux upper-limit above ${\sim}3$~EeV --- while simultaneously paving the way for future large-array UHE neutrino observatories.\\

% Bibtex references:
\begingroup
\setstretch{0.5}
\setlength{\bibsep}{1.0pt}
\bibliographystyle{ICRC}
\bibliography{references}
\endgroup

% Alternatively, you can include references by hand:
%\begin{thebibliography}{99}
%\bibitem{...}
%
%\end{thebibliography}

\clearpage

%The following list of authors, affiliations and funding agencies should be updated at the day of submission. 
% ICRC list for ARA Collaboration
\section*{Full Author List: ARA Collaboration (June 30, 2025)}

\noindent
N.~Alden\textsuperscript{1}, 
S.~Ali\textsuperscript{2}, 
P.~Allison\textsuperscript{3}, 
S.~Archambault\textsuperscript{4}, 
J.J.~Beatty\textsuperscript{3}, 
D.Z.~Besson\textsuperscript{2}, 
A.~Bishop\textsuperscript{5}, 
P.~Chen\textsuperscript{6}, 
Y.C.~Chen\textsuperscript{6}, 
Y.-C.~Chen\textsuperscript{6}, 
S.~Chiche\textsuperscript{7}, 
B.A.~Clark\textsuperscript{8}, 
A.~Connolly\textsuperscript{3}, 
K.~Couberly\textsuperscript{2}, 
L.~Cremonesi\textsuperscript{9}, 
A.~Cummings\textsuperscript{10,11,12}, 
P.~Dasgupta\textsuperscript{3}, 
R.~Debolt\textsuperscript{3}, 
S.~de~Kockere\textsuperscript{13}, 
K.D.~de~Vries\textsuperscript{13}, 
C.~Deaconu\textsuperscript{1}, 
M.A.~DuVernois\textsuperscript{5}, 
J.~Flaherty\textsuperscript{3}, 
E.~Friedman\textsuperscript{8}, 
R.~Gaior\textsuperscript{4}, 
P.~Giri\textsuperscript{14}, 
J.~Hanson\textsuperscript{15}, 
N.~Harty\textsuperscript{16}, 
K.D.~Hoffman\textsuperscript{8}, 
M.-H.~Huang\textsuperscript{6,17}, 
K.~Hughes\textsuperscript{3}, 
A.~Ishihara\textsuperscript{4}, 
A.~Karle\textsuperscript{5}, 
J.L.~Kelley\textsuperscript{5}, 
K.-C.~Kim\textsuperscript{8}, 
M.-C.~Kim\textsuperscript{4}, 
I.~Kravchenko\textsuperscript{14}, 
R.~Krebs\textsuperscript{10,11}, 
C.Y.~Kuo\textsuperscript{6}, 
K.~Kurusu\textsuperscript{4}, 
U.A.~Latif\textsuperscript{13}, 
C.H.~Liu\textsuperscript{14}, 
T.C.~Liu\textsuperscript{6,18}, 
W.~Luszczak\textsuperscript{3}, 
A.~Machtay\textsuperscript{3}, 
K.~Mase\textsuperscript{4}, 
M.S.~Muzio\textsuperscript{5,10,11,12}, 
J.~Nam\textsuperscript{6}, 
R.J.~Nichol\textsuperscript{9}, 
A.~Novikov\textsuperscript{16}, 
A.~Nozdrina\textsuperscript{3}, 
E.~Oberla\textsuperscript{1}, 
C.W.~Pai\textsuperscript{6}, 
Y.~Pan\textsuperscript{16}, 
C.~Pfendner\textsuperscript{19}, 
N.~Punsuebsay\textsuperscript{16}, 
J.~Roth\textsuperscript{16}, 
A.~Salcedo-Gomez\textsuperscript{3}, 
D.~Seckel\textsuperscript{16}, 
M.F.H.~Seikh\textsuperscript{2}, 
Y.-S.~Shiao\textsuperscript{6,20}, 
S.C.~Su\textsuperscript{6}, 
S.~Toscano\textsuperscript{7}, 
J.~Torres\textsuperscript{3}, 
J.~Touart\textsuperscript{8}, 
N.~van~Eijndhoven\textsuperscript{13}, 
A.~Vieregg\textsuperscript{1}, 
M.~Vilarino~Fostier\textsuperscript{7}, 
M.-Z.~Wang\textsuperscript{6}, 
S.-H.~Wang\textsuperscript{6}, 
P.~Windischhofer\textsuperscript{1}, 
S.A.~Wissel\textsuperscript{10,11,12}, 
C.~Xie\textsuperscript{9}, 
S.~Yoshida\textsuperscript{4}, 
R.~Young\textsuperscript{2}
\\
\\
\textsuperscript{1} Dept. of Physics, Enrico Fermi Institute, Kavli Institute for Cosmological Physics, University of Chicago, Chicago, IL 60637\\
\textsuperscript{2} Dept. of Physics and Astronomy, University of Kansas, Lawrence, KS 66045\\
\textsuperscript{3} Dept. of Physics, Center for Cosmology and AstroParticle Physics, The Ohio State University, Columbus, OH 43210\\
\textsuperscript{4} Dept. of Physics, Chiba University, Chiba, Japan\\
\textsuperscript{5} Dept. of Physics, University of Wisconsin-Madison, Madison,  WI 53706\\
\textsuperscript{6} Dept. of Physics, Grad. Inst. of Astrophys., Leung Center for Cosmology and Particle Astrophysics, National Taiwan University, Taipei, Taiwan\\
\textsuperscript{7} Universite Libre de Bruxelles, Science Faculty CP230, B-1050 Brussels, Belgium\\
\textsuperscript{8} Dept. of Physics, University of Maryland, College Park, MD 20742\\
\textsuperscript{9} Dept. of Physics and Astronomy, University College London, London, United Kingdom\\
\textsuperscript{10} Center for Multi-Messenger Astrophysics, Institute for Gravitation and the Cosmos, Pennsylvania State University, University Park, PA 16802\\
\textsuperscript{11} Dept. of Physics, Pennsylvania State University, University Park, PA 16802\\
\textsuperscript{12} Dept. of Astronomy and Astrophysics, Pennsylvania State University, University Park, PA 16802\\
\textsuperscript{13} Vrije Universiteit Brussel, Brussels, Belgium\\
\textsuperscript{14} Dept. of Physics and Astronomy, University of Nebraska, Lincoln, Nebraska 68588\\
\textsuperscript{15} Dept. Physics and Astronomy, Whittier College, Whittier, CA 90602\\
\textsuperscript{16} Dept. of Physics, University of Delaware, Newark, DE 19716\\
\textsuperscript{17} Dept. of Energy Engineering, National United University, Miaoli, Taiwan\\
\textsuperscript{18} Dept. of Applied Physics, National Pingtung University, Pingtung City, Pingtung County 900393, Taiwan\\
\textsuperscript{19} Dept. of Physics and Astronomy, Denison University, Granville, Ohio 43023\\
\textsuperscript{20} National Nano Device Laboratories, Hsinchu 300, Taiwan\\

\section*{Acknowledgements}

\noindent
This analysis was conducted by Abigail Bishop, Brian Clark, Paramita Dasgupta, Pawan Giri, Alan Salcedo Gomez, Alex Machtay, Marco Muzio, and Mohammad Ful Hossain Seikh.
The ARA Collaboration is grateful to support from the National Science Foundation through Award 2013134.
The ARA Collaboration
designed, constructed, and now operates the ARA detectors. We would like to thank IceCube, and specifically the winterovers for the support in operating the
detector. Data processing and calibration, Monte Carlo
simulations of the detector and of theoretical models
and data analyses were performed by a large number
of collaboration members, who also discussed and approved the scientific results presented here. We are
thankful to Antarctic Support Contractor staff, a Leidos unit 
for field support and enabling our work on the harshest continent. We thank the National Science Foundation (NSF) Office of Polar Programs and
Physics Division for funding support. We further thank
the Taiwan National Science Councils Vanguard Program NSC 92-2628-M-002-09 and the Belgian F.R.S.-
FNRS Grant 4.4508.01 and FWO. 
K. Hughes thanks the NSF for
support through the Graduate Research Fellowship Program Award DGE-1746045. A. Connolly thanks the NSF for
Award 1806923 and 2209588, and also acknowledges the Ohio Supercomputer Center. S. A. Wissel thanks the NSF for support through CAREER Award 2033500.
A. Vieregg thanks the Sloan Foundation and the Research Corporation for Science Advancement, the Research Computing Center and the Kavli Institute for Cosmological Physics at the University of Chicago for the resources they provided. R. Nichol thanks the Leverhulme
Trust for their support. K.D. de Vries is supported by
European Research Council under the European Unions
Horizon research and innovation program (grant agreement 763 No 805486). D. Besson, I. Kravchenko, and D. Seckel thank the NSF for support through the IceCube EPSCoR Initiative (Award ID 2019597). M.S. Muzio thanks the NSF for support through the MPS-Ascend Postdoctoral Fellowship under Award 2138121. A. Bishop thanks the Belgian American Education Foundation for their Graduate Fellowship support.

\end{document}